\pdfoutput=1
\documentclass[aps,pre,reprint,groupedaddress,longbibliography]{revtex4-1}	

\usepackage{amssymb}
\usepackage{color}
\usepackage{amsmath}
\usepackage{grffile}
\usepackage{cancel}
\usepackage{hyperref}

\hypersetup{
  pdfproducer= none,  
  pdfcreator = none,  
  colorlinks = true,
  allcolors = black, 
}

\usepackage{tabularx, booktabs}
\newcolumntype{Y}{>{\centering\arraybackslash}X}

\usepackage[usenames,dvipsnames]{xcolor}

\usepackage{rotating}

\usepackage[english]{babel}

\usepackage{graphicx}

\newcommand{\Fig}[1]{Fig.~\ref{#1}}

\newcommand{\Tab}[1]{Table~\ref{#1}}

\newcommand{\imagedir}{./}

\usepackage{array}
\newcolumntype{C}[1]{>{\centering\arraybackslash}p{#1}}

\def\eqq#1{Eq.~(\ref{#1})}
\def\eq#1{(\ref{#1})}

\def\f#1{Fig.~\ref{#1}}

\def\c#1{~\cite{#1}}
\def\cc#1{Ref.~\cite{#1}}

\def\beq{\begin{equation}}
\def\eeq{\end{equation}}
\def\bea{\begin{eqnarray}}
\def\eea{\end{eqnarray}}

\def\fb{f_{\rm b}}
\def\fbs{f_{\rm b}^{\rm s}}

\def\kt{k_{\rm B}T}

\begin{document}

\title{Predicting the outcome of the growth of binary solids far from equilibrium}

\author{Ranjan V. Mannige}
\email[]{rvmannige@lbl.gov}
\affiliation{Molecular Foundry, Lawrence Berkeley National Laboratory, 1 Cyclotron Road, Berkeley, CA, U.S.A.}
\author{Stephen Whitelam}
\email[]{swhitelam@lbl.gov}
\affiliation{Molecular Foundry, Lawrence Berkeley National Laboratory, 1 Cyclotron Road, Berkeley, CA, U.S.A.}

\date{\today}

\begin{abstract}
The growth of multicomponent structures in simulations and experiments often results in kinetically trapped, nonequilibrium objects. In such cases we have no general theoretical framework for predicting the outcome of the growth process. Here we use computer simulations to study the growth of two-component structures within a simple lattice model. We show that kinetic trapping happens for many choices of growth rate and inter-component interaction energies, and that qualitatively distinct kinds of kinetic trapping are found in different regions of parameter space. In a region in which the low-energy structure is an `antiferromagnet' or  `checkerboard', we show that the grown nonequilibrium structure displays a component-type stoichiometry that is different to the equilibrium one but is insensitive to growth rate and solution conditions. This robust nonequilibrium stoichiometry can be predicted via a mapping to the jammed random tiling of dimers studied by Flory, a finding that suggests a way of making defined nonequilibrium structures in experiment. 
\end{abstract}


\maketitle

\section{Introduction} 

Molecular self-assembly is the spontaneous organization of components, which move around under e.g. Brownian motion but are otherwise left undisturbed, into ordered structures. Self-assembly holds considerable promise for materials science\c{whitesides2002self,glotzer2007anisotropy,frenkel2015order,cademartiri2015programmable}. The goal of molecular self-assembly in the laboratory is often to make an equilibrium structure, and the laws of statistical mechanics indeed dictate that components undergoing Brownian motion will eventually build themselves into the 
structure of least free energy. In practical terms, however, slow dynamical processes can prevent equilibration from happening on the timescale available to the process in question\c{doi:10.1146/annurev-physchem-040214-121215}. In such circumstances the processes of nucleation and growth lead instead to the formation of kinetically trapped, nonequilibrium structures. Multicomponent systems, i.e. systems composed of more than one type of component, are particularly susceptible to kinetic trapping because the slow rearrangement of component types within a solid structure can prevent them from achieving their equilibrium arrangement as the solid structure grows. Frequently, the outcome of the nucleation and growth of multiple component types is an ordered crystal structure within which component types are arranged in a nonequilibrium way\c{Sanz2007,Peters2009,Kim2009,Scarlett2010,Whitelam2014a}. Such structures have potentially useful properties. However, predicting their component-type arrangements is not possible in general, because we cannot predict the outcome of self-assembly when that outcome is not the equilibrium structure.

Here we use simulation and analytic theory to study the component-type arrangements formed during the growth of two-component structures within a simple lattice model. In accord with several experimental results, growth results in the formation of nonequilibrium structures for a large range of growth rates and inter-component interaction energies. In some regions of parameter space the properties of nonequilibrium structures vary continuously with growth rate, while in other regimes of parameter space these properties are insensitive to growth rate. In a region of parameter space in which the low-energy structure is a binary `checkerboard' we show that the grown nonequilibrium structure displays a component-type stoichiometry that is insensitive both to growth rate and to the abundance of component types in solution. We show that this robust nonequilibrium stoichiometry can be predicted via a mapping to the jammed random tiling of dimers studied by Flory. These findings suggest a route to the rational design of defined nonequilibrium structures in experiment.

\section{Model and Simulation Methods}
\begin{figure}[t!]
 \includegraphics[width=0.45\textwidth]{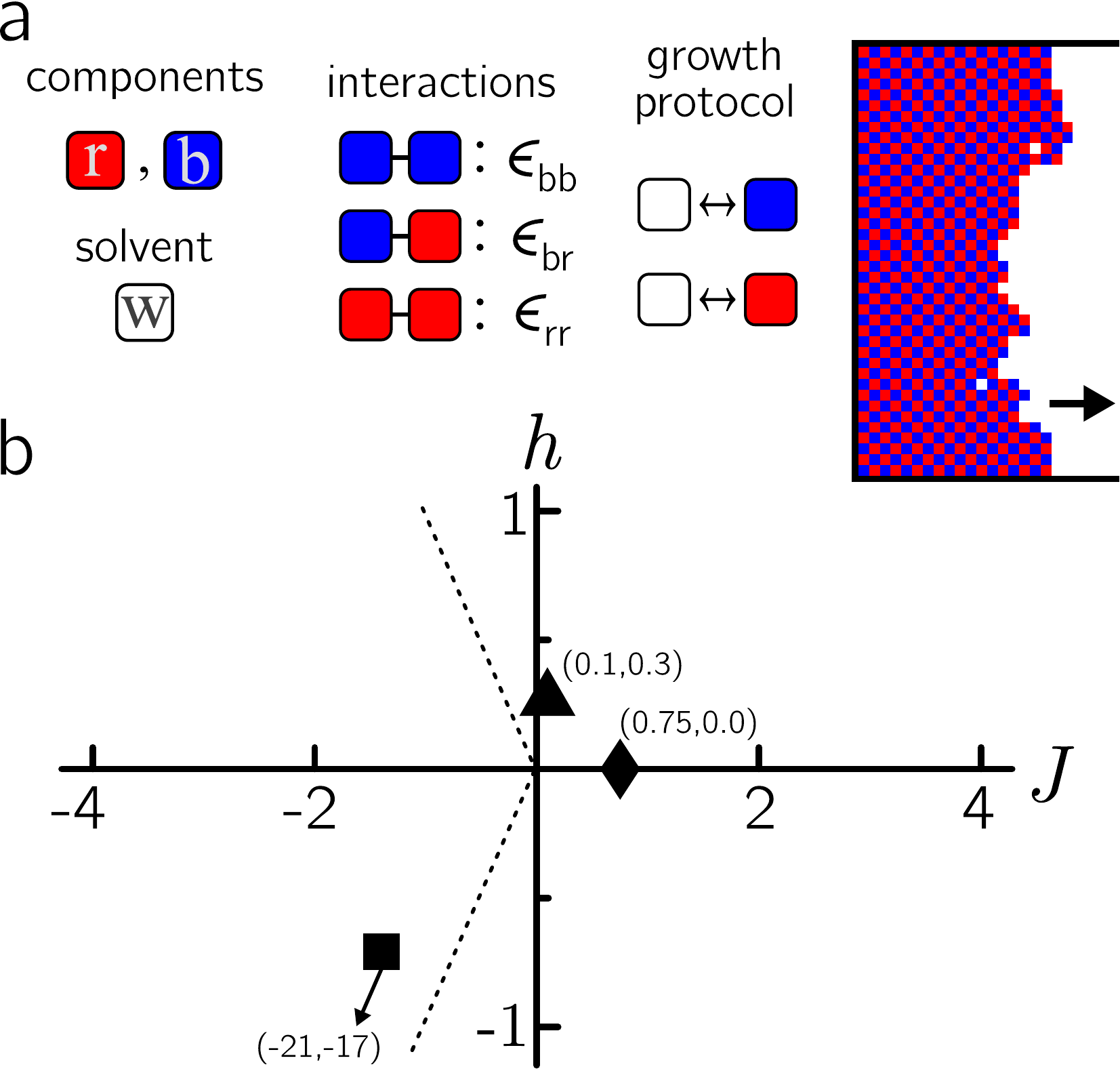}
  \caption{(a) Schematic of the lattice model and Monte Carlo protocol we use in this paper to study growth. (b) Distinct kinds of kinetic trapping can be found for different combinations of red-blue interaction energies (see definitions of $J$ and $h$ in the text). In this paper we focus on the region of phase space to the left of the dotted line, where the low-energy structure is a red-blue checkerboard. We also comment on growth at points $\blacksquare$\c{Sue2015}, $\blacktriangle$\c{Kim2009}, and $\blacklozenge$\c{Whitelam2014a}, considered in previous studies.
\label{FigGrowthProtocol}}
\end{figure}

Kinetic trapping of component types within growing multicomponent structures has a simple physical origin -- the slow dynamics of particles within a solid -- and so can be reproduced by simple physical models that account for this slow dynamics\c{Sanz2007,Peters2009,Kim2009,Scarlett2010}. Here we consider a lattice model of growth similar to the models used in Refs.\c{Whitelam2014a,Hedges2014,Sue2015}. We focus on growth in a 2D system, but we will also present results for higher dimensions. As sketched in \Fig{FigGrowthProtocol}(a), lattice sites can be unoccupied (white) or occupied by a particle of one of two types (red or blue). Red and blue particles (or components)
experience color-dependent nearest-neighbor interactions (see Appendix~\ref{app}) of energy $\epsilon_{\rm rr}$, $\epsilon_{\rm br}$, and $\epsilon_{\rm bb}$, in units of $k_{\rm B}T$ (which we shall set equal to unity). On a fully-occupied lattice (one without white sites) these interactions are equivalent to the Ising model with magnetic field $h \equiv (\epsilon_\textrm{rr} - \epsilon_\textrm{bb})/4$ and coupling constant $J \equiv \epsilon_\textrm{br}/2 - (\epsilon_\textrm{rr} + \epsilon_\textrm{bb})/4$\c{Sue2015}. White, blue, and red sites also receive energetic penalties $\mu$, $-\ln \fbs$ and $-\ln (1-\fbs)$, respectively. Here $\mu$ sets the relative abundance of colored and white sites in notional `solution' (i.e. in the absence of energetic interactions), and $\fbs$ is the notional solution fraction of colored blocks that are blue. We evolved this model using a grand-canonical Monte Carlo procedure that respects detailed balance, and that resolves the stochastic binding and unbinding of red and blue components. Unbinding dynamics is naturally slow when components possess many colored neighbors; we also imposed a kinetic constraint that prevents any change of state of a lattice site that possesses only colored neighbors. This constraint, which preserves detailed balance, is intended to model the fact that relaxation dynamics within solid structures is slow. In what follows we shall describe growth simulations done in the presence and absence of the kinetic constraint. The latter type of simulation represents a convenient way to assess the outcome of growth on timescales longer than we could otherwise access. In most simulations described below we used a 2D square lattice of $40\times400$ lattice sites whose long edges were periodic and whose short edges were not. We began simulations in the presence of a `seed' at the left-hand short edge of the simulation box, with the rest of the box left white, so that we could study growth without waiting for nucleation to happen. By varying $\mu$ we could change the rate of growth of the colored assembly. In what follows we refer to `growth' simulations in which the simulation was stopped after 90\% of the box become occupied by colored components, and `maturation' simulations in which structures grown in this manner were allowed to evolve for an additional $10^3 - 10^5$ Monte Carlo cycles.

\section{Growth simulations} 
\begin{figure*}[t!]
\centering
\includegraphics[width=0.95\textwidth]{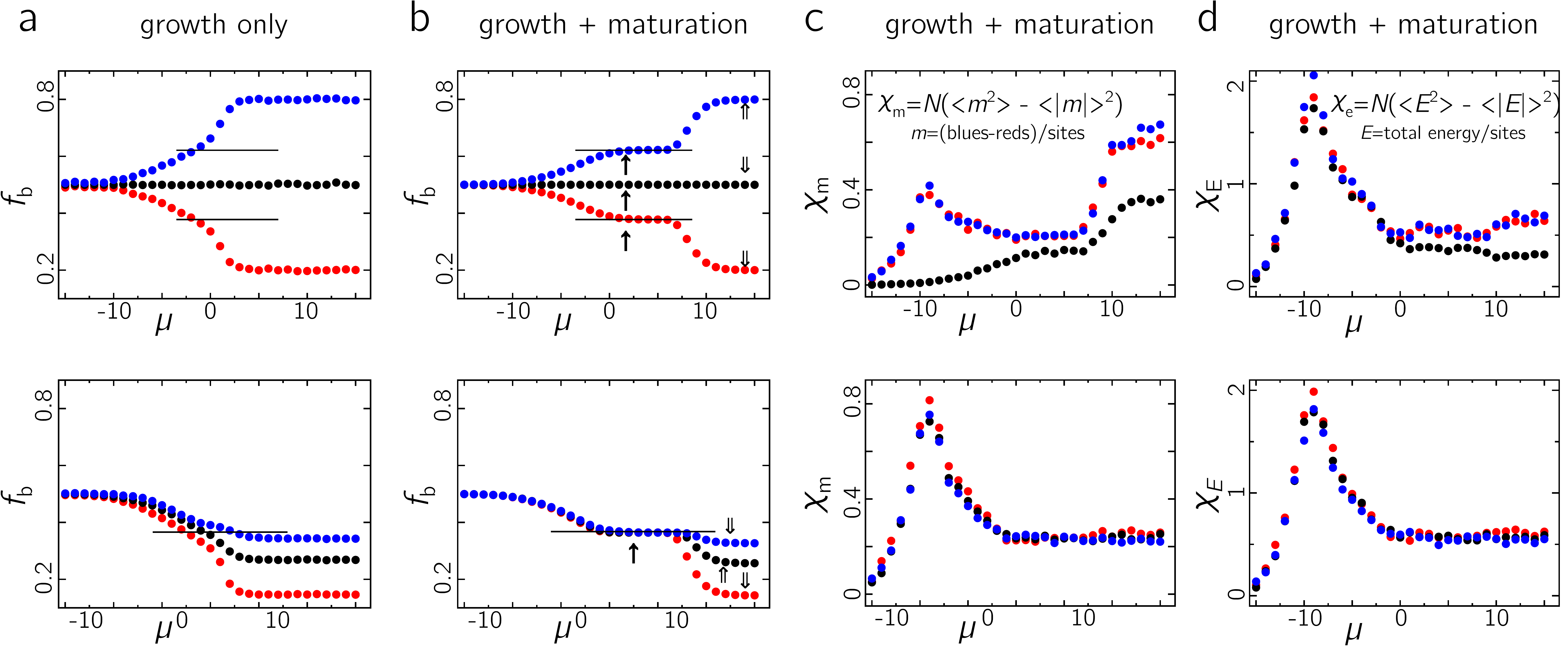} 
  \caption{The outcome of (a) growth and (b) growth-and-maturation simulations for symmetric (top panel: $\epsilon_\textrm{br}<0\equiv\epsilon_\textrm{rr}\equiv\epsilon_\textrm{bb}$) and asymmetric (bottom panel: $\epsilon_\textrm{br}<0\equiv\epsilon_\textrm{rr}\ll\epsilon_\textrm{bb}$) interaction energy hierarchies reveals the existence of `mature' nonequilibrium structures whose stoichiometry is insensitive to growth rate (b, top panel) and growth rate and solution stoichiometry (b, bottom panel). Here $\fb$ is the fraction of colored components in the grown structures that are blue, and $\mu$ is a chemical potential: the larger is $\mu$, the more rapid is the rate of growth. Growth simulations were done using three distinct solution fractions of blue components, $\fbs=$ 0.2, 0.5, and 0.8 (red, black, and blue lines). Panels (c) and (d) show that near-equilibrium and far-from-equilibrium regimes are separated by a regime of large fluctuations of color and energy (measured using $10^3$ independent simulations at each value of $\mu$), suggesting the presence of a nonequilibrium phase transition.
\label{FigFbs}}
\end{figure*}

Growth carried out using different choices of the inter-component energetic parameters $J$ and $h$~\footnote{The Ising parameters $J$ and $h$ are a convenient way to describe the three blue-red interactions $\epsilon_{\rm rr}$, $\epsilon_{\rm br}$, and $\epsilon_{\rm bb}$, but they provide only a partial description of the model's parameter space, which includes chemical potential terms and interactions with white sites. For instance, one can add constant terms to the $\epsilon$ parameters that leave $J$ and $h$ unchanged but change the energetics of the model.}, shown in \Fig{FigGrowthProtocol}(b), is similar in the following respects (see \Fig{FigFbs}). At vanishing rates of growth a structure resembling the equilibrium one is generated; at very large rates of growth a `solid solution' is obtained, i.e. red and blue components are arranged randomly on the lattice in proportion to their solution proportions; and at intermediate rates of growth one obtains nonequilibrium structures that differ from both of these limiting cases. These nontrivial nonequilibrium structures can be different in different parameter regimes. For instance, using the `ferromagnetic' energy scale hierarchy shown by the symbol $\blacklozenge$ on \Fig{FigGrowthProtocol}(b), nonequilibrium structures include `critical' arrangements in which red and blue component domains of a broad size distribution are present\c{Whitelam2014a} (this behavior may be related to that seen in certain irreversible cellular automata\c{ausloos1993magnetic,candia2008magnetic}). At the parameter combination labeled $\blacktriangle$, nonequilibrium structures consist of large domains of the blue component within which a small red impurity fraction is found (see \Fig{Fig:Crocker2}). This impurity fraction is only weakly sensitive to growth rate over some range of growth rates, a result that reproduces the qualitative outcome of growth in experiments and off-lattice simulations done by other authors\c{Kim2009}. The reproduction of these results by the present model suggests that it captures key physical aspects of real growth processes. Finally, at the parameter combination labeled $\blacksquare$ on \Fig{FigGrowthProtocol}(b), growth results in a nonequilibrium component-type stoichiometry identical (on an {\bf nbo} lattice) to that seen in a certain metal-organic framework; this stoichiometry is insensitive to growth rate and component solution stoichiometry over some range of those parameters\c{Sue2015}. As we shall show, robust nonequilibrium stoichiometry is also seen in other parameter regimes left of the dotted line in \f{FigGrowthProtocol}. Here we aim to provide a partial physical understanding of this behavior.

We start by showing in \Fig{FigFbs} the outcome of growth simulations done in the aforementioned regime of parameter space, where the red-blue energetic interaction is lower in energy than both of the like-color interactions. Here the low-energy structure, and the thermodynamically stable structure for the parameter values we shall consider, is an alternating red-blue `antiferromagnet' or `checkerboard'. In the top panel of \Fig{FigFbs} we show results for the parameter combination $h=0$ (meaning that red-red and blue-blue interactions are of equal strength), while in the bottom panel we consider an asymmetric energy hierarchy for which $h\neq0$. At low rates of growth, in both cases, the structure generated dynamically, upon 95\% filling of the simulation box (panel (a)), is close in nature to the equilibrium structure, and so possesses a blue fraction (fraction of colored components that are blue) $\fb= 1/2$. For large rates of growth the structures obtained are kinetically trapped arrangements of components whose blue fraction is related to that of the notional solution (we considered three different solution stochiometries, shown as black, red, and blue lines). At intermediate rates of growth the blue fraction of the grown structure interpolates smoothly between these limiting cases. Configuration snapshots are shown in \f{FigGrowthVsMaturationSnapshotsAll}. However, when allowed to further evolve or `mature' for $10^4$ Monte Carlo sweeps (panel (b)) (in the absence of the kinetic constraint so as to effectively allow access to longer timescales), structures generated at intermediate growth rates did not evolve to equilibrium, but instead became kinetically trapped in configurations whose blue fractions display plateaux as a function of growth rate. That is, the stoichiometry of those nonequilibrium structure is insensitive to growth rate. Furthermore, in the case of the asymmetric energetic hierarchy (bottom panel) this stoichiometry was {\em also} insensitive to solution stoichiometry. Near-equilibrium and far-from-equilibrium regimes are separated by a regime of large fluctuations of color and energy (panels (c) and (d)), suggesting the existence of a nonequilibrium phase transition similar to that seen in the `ferromagnetic' regime of parameter space\c{Whitelam2014a}.

Structures generated dynamically in the presence of certain energetic interactions therefore display a stoichiometry that is different to the 1:1 equilibrium one, but that is robust with respect to changes of growth rate and solution stoichiometry over a considerable range of those parameters. We call these robust nonequilibrium stochiometries `magic numbers'. The existence of magic numbers has potential application for materials science, because it suggests that one can grow two-component solids, out of equilibrium, in a predictable manner. Magic number materials may have already been synthesized. One particular two-component metal-organic framework (MOF), called MOF-2000, displays a stoichiometry that is robust to solution stoichiometry over a considerable range\c{Sue2015}. The numerical value of this stoichiometry can be reproduced by the growth of magic-number structures using the present model on a 3D framework whose topology is appropriate to the crystal structure of MOF-2000\c{Sue2015}. 

\begin{figure}[t!]
  \centering
  \includegraphics[width=0.35\textwidth]{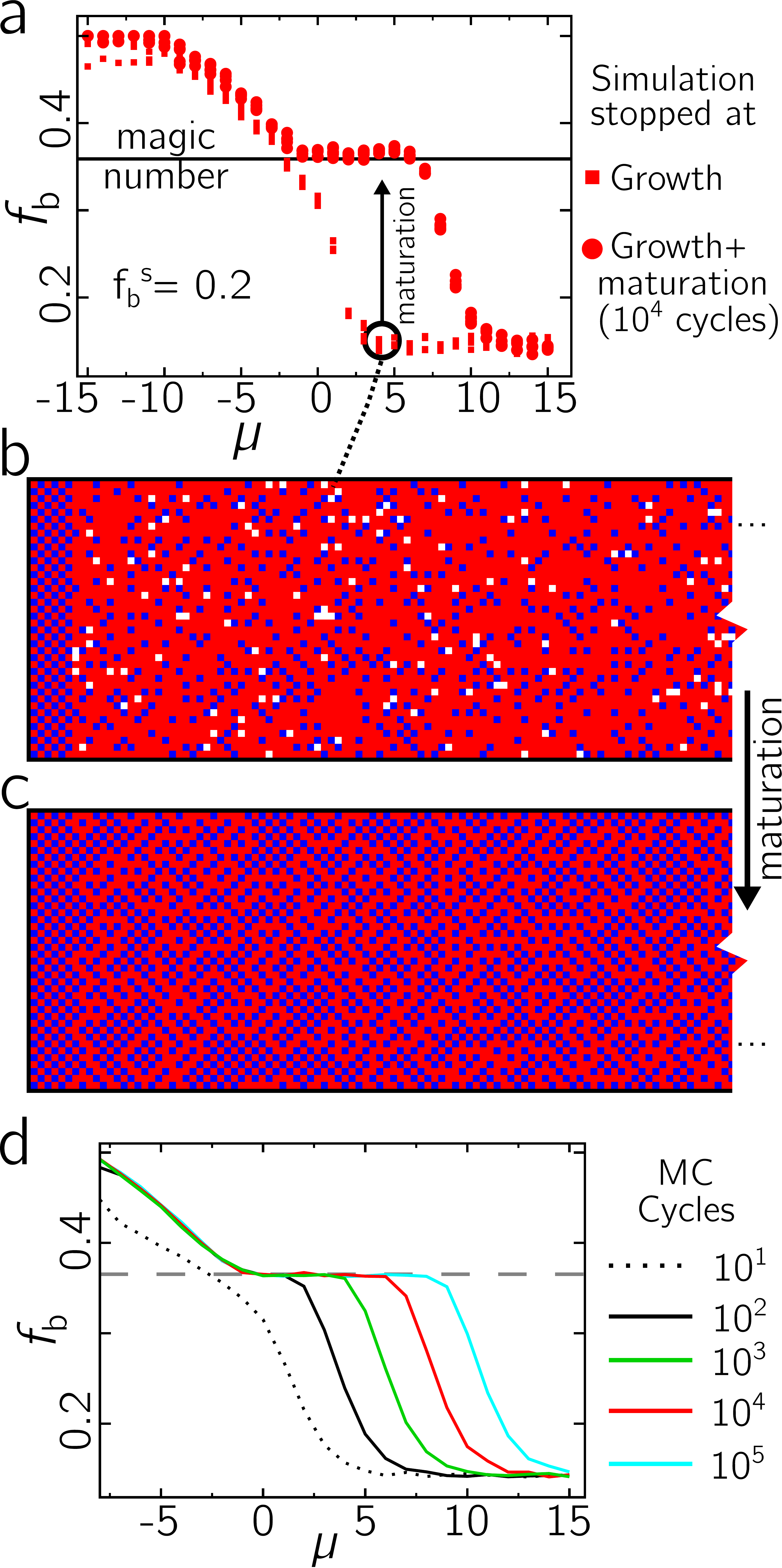}
  \caption{Fast growth followed by maturation results in `magic number' structures. (a) The blue fraction $\fb$ for freshly-grown structures varies smoothly with growth rate between equilibrium and far-from-equilibrium limits. If allowed to evolve further, structures grown at a range of rates evolve to nonequilibrium structures that possess the same `magic number' stoichiometry. (b) and (c) show `grown' and `mature' structures corresponding to the points indicated on the top panel. See also \f{FigGrowthVsMaturationSnapshotsAll} and \f{MoreConditions}. Maturation was stopped at $10^4$ Monte Carlo cycles in (a); stopping the simulations between $10^2$ and $10^5$ cycles yield similar plateaux (d; \f{ManyCycles}). 
\label{FigGrowthVsMaturation}}
\end{figure}

In physical terms, nonequilibrium structures emerge because microscopic contacts that are not the equilibrium or `native' red-blue one (such as red-red contacts) can appear stochastically during growth and can become trapped by the arrival of additional material. In some regions of parameter space the properties of the resulting kinetically trapped structures vary continuously with growth rate and solution stoichiometry; in the magic number regime they do not. As shown in \Fig{FigGrowthVsMaturation}, magic number configurations, reached upon `maturation' of a structure after its initial growth, are long-lived: 
evolution to the equilibrium checkerboard structure does not happen on the timescale of simulation (\f{FigGrowthVsMaturation}d; \f{ManyCycles}). 
Magic number structures can also be generated by zero-temperature, single-spin-flip Monte Carlo sampling of fully-occupied lattices; that is, magic number structures are also inherent structures of the lattice Hamiltonian. These inherent structures are accessible from a wide range of initial conditions (they have a large basin of attraction), and the numerical values of the associated magic numbers are dependent upon lattice connectivity and dimensionality (\Fig{Fig0}).

\begin{figure}[t!]
  \centering
  \includegraphics[width=0.5\textwidth]{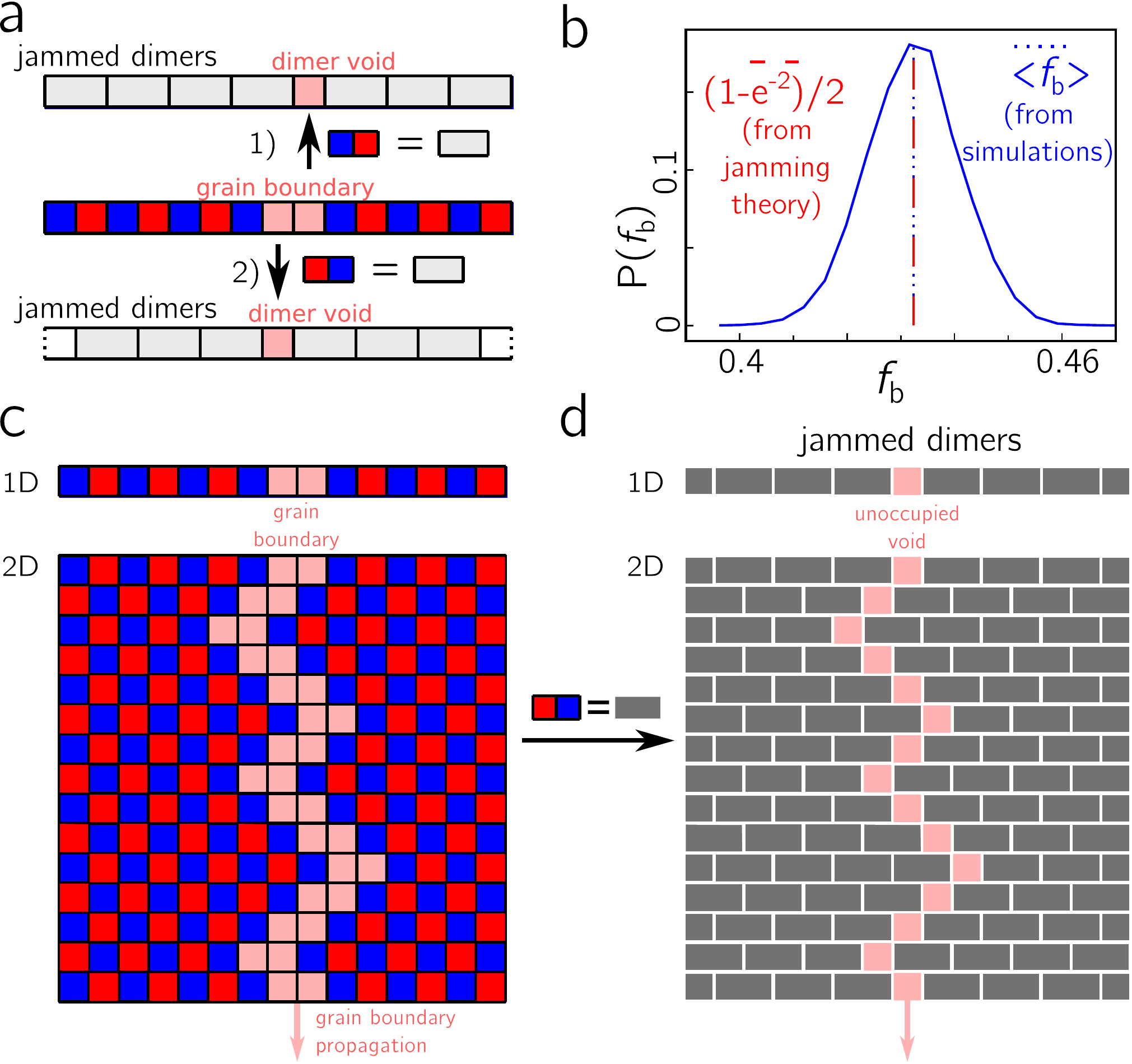}
  \caption{(a,b) In $d=1$ the inherent structures of the lattice model with no white sites are, for the energetic hierarchy $\epsilon_\textrm{br}<0\equiv\epsilon_\textrm{rr}\ll\epsilon_\textrm{bb}$, equivalent to those produced by random sequential absorption (RSA) of dimers on a lattice. (c,d) This equivalence does not hold in higher dimensions, but there we can nonetheless use the jamming result to approximate the inherent structure result via the graphical construction shown (see text). The resulting prediction, \eqq{flory_ext}, is in reasonable accord with inherent structure results for $d \leq 3$ (see \f{fig_dimension}). Because the long-time outcome of growth simulations for this energetic hierarchy are inherent structures of the lattice model with no white sites, the same magic numbers are seen in our growth simulations, i.e. in \f{FigFbs}(b) bottom and \f{FigGrowthVsMaturation}. Thus, the nonequilibrium stoichiometry resulting from growth can be predicted via a mapping to a jammed system of dimers.  
\label{fig_flory}}
\end{figure}

\section{Mapping to jammed dimer systems}

The interaction energies used to obtain the magic numbers seen in \f{FigFbs} satisfy the hierarchy $\epsilon_\textrm{br}<\epsilon_\textrm{rr}\ll\epsilon_\textrm{bb}$. In words, the blue-red contact is the `native' or equilibrium one; red-red contacts are higher in energy but can occur during growth; and blue-blue contacts are so unfavorable that they cannot form at reasonable rates of growth. In one dimension this energetic hierarchy results in the growth of red-blue arrangements, such as those shown in \f{fig_flory}(a), that map to a tiling of dimers with voids. `Dimers' are red-blue pairs, and `voids' are red particles. The particle-to-dimer mapping produces one of two equivalent void arrangements, as shown. The long-time outcome of our growth process then becomes equivalent to that of random sequential absorption (RSA) of dimers on a one-dimensional lattice. This problem was studied by Flory\c{Flory1939}, who computed the dimer filling fraction to be $1-{\rm e}^{-2}$. The mean blue fraction of our equivalent red-blue structure is then half of this value, i.e. $\fb = (1-{\rm e}^{-2})/2 \approx 0.432$. This value is indeed the mean value of the stoichiometry of inherent structures of our lattice model (recall that in this regime of parameter space the `matured' growth configurations are also inherent structures of the model with no white sites), which we shown in \f{fig_flory}(b). Thus, the nonequilibrium stoichiometry resulting from growth can be predicted via a mapping to a jammed system of dimers.  

\begin{figure}[tb]
  \centering
  \includegraphics[width=0.44\textwidth]{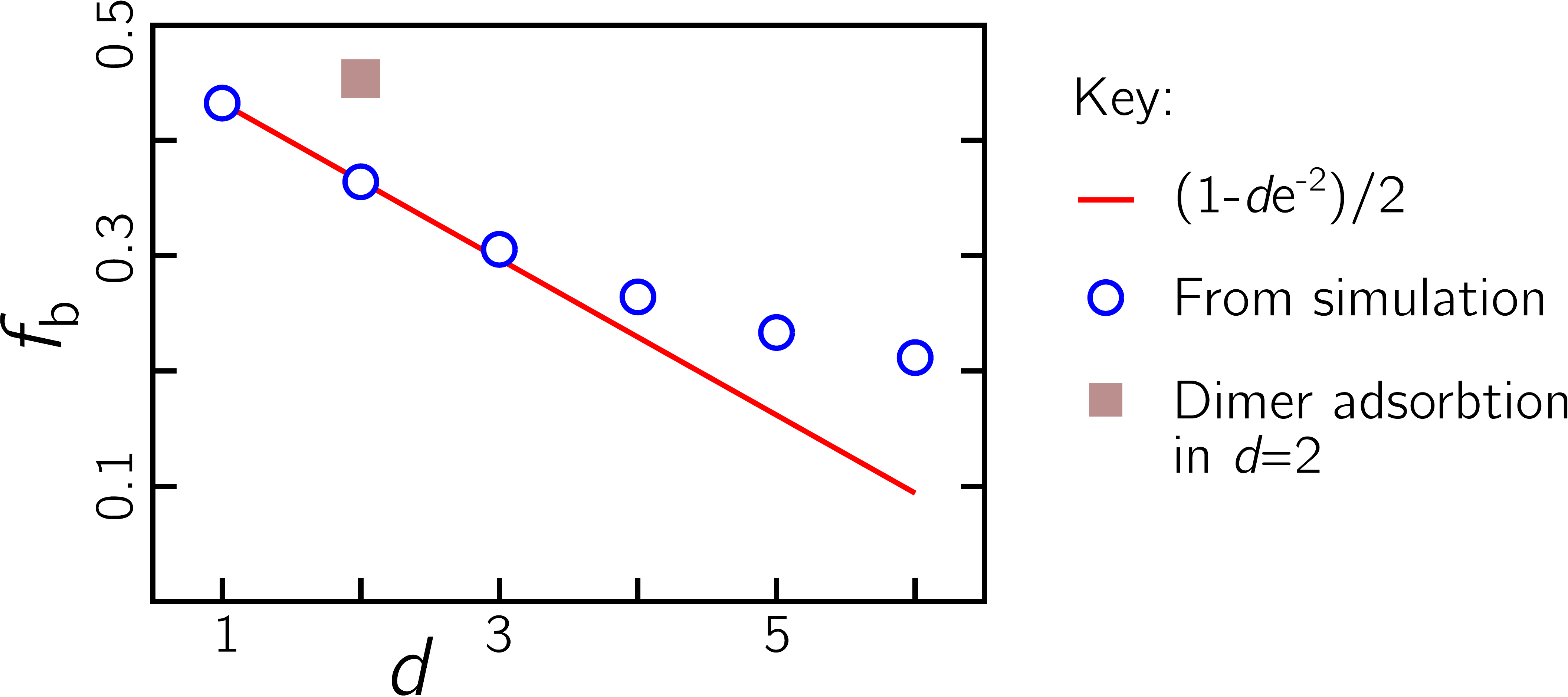}
  \caption{Our approximate extrapolation of Flory's dimer-packing result, \eqq{flory_ext} (red line), matches with reasonable precision the nonequilibrium `magic number' stochiometries seen in inherent structures of the lattice model in $d \leq 3$ dimensions when component interactions satisfy the hierarchy $\epsilon_\textrm{br}<\epsilon_\textrm{rr}\ll\epsilon_\textrm{bb}$. The plateux seen in growth simulations in \f{FigFbs}(b, bottom) and \f{FigGrowthVsMaturation} have numerical values similar to the point at $d=2$ here. For dimensions $d \geq 4$ the analytic and numerical results deviate. 
\label{fig_dimension}}
\end{figure}

The equivalence between our growth process and dimer deposition does not hold in dimensions greater than one. Nonetheless, we can use the Flory result to estimate numerically the magic number ratio seen in our growth simulations in 2D and 3D. Consider a periodic hypercubic lattice that possesses $N$ lattice sites or nodes in each dimension, and so has $N^d$ nodes in total. Each node may be occupied by one red or one blue particle. Let $V$ be the number of void sites that exist on a connected row of $N$ nodes in any given dimension, and assume that $V/N={\rm e^{-2}}$\c{Flory1939}. Assume further that each dimension is independent, so that each void site connects to a continuous chain of void sites that extends independently into each of the remaining $d-1$ dimensions; see \f{fig_flory}(c,d). Thus, each void region contains in total $V N^{d-1}$ voids. Summed over all independent dimensions there therefore exist $d V N^{d-1}$ voids in total, meaning that the void density is $d V/N = d {\rm e^{-2}}$. We therefore predict the nonequilibrium `magic number' stoichiometry of our red-blue structure, grown in $d$ dimensions, to be 
\beq
\label{flory_ext}
\fb^d=(1-d{\rm e^{-2}})/2.
\eeq
The magic number structures seen in the 2D growth processes whose results are reported in \f{FigFbs}(b, bottom) and \f{FigGrowthVsMaturation} have a stoichiometry (magic number blue fraction) of 
$0.364\pm0.0035$.
The estimate of \eqq{flory_ext} is $f_b^2 = (1-2 {\rm e}^{-2})/2 \approx 0.365$, which agrees closely with our inherent simulation results (\f{fig_dimension}) and with the plateaux seen in our growth simulations (\f{FigGrowthVsMaturation} and \f{FigFbs}(b) bottom). Thus the expression \eq{flory_ext} can be used to predict the nature of a kinetically trapped structure generated in 2D by two-component growth. In 3D the estimate \eq{flory_ext} predicts a magic number blue fraction of $\fb^3=0.297$; our inherent structure simulations done in 3D display a very similar magic number ratio of 
$0.3\pm0.0026$;
see \Fig{fig_dimension}. In dimension 4 and up the predictions of \eqq{flory_ext} depart from the results of our numerical simulations (see \Fig{fig_dimension}), signaling the breakdown of the approximations we used to derive the equation. But in dimensions relevant to laboratory self-assembly, for square and cubic crystal structures, we can rationalize the stoichiometry that results from two-component kinetic trapping by analogy to a jamming problem.

We have shown that of a lattice model of two-component growth displays a rich range of phenomenology, key aspects of which reproduce behavior seen in experiments\c{Kim2009,Sue2015}. Growth can result in near-equilibrium structures and far-from-equilibrium structures. In certain regimes of parameter space the component-type stoichiometry of these nonequilibrium structures is independent of growth rate and solution stoichiometry, and the numerical value of this stoichiometry can be predicted via a mapping to a jammed tiling of dimers. These observations suggest that one can grow, far from equilibrium, defined two-component structures in experiment. 

\appendix 
\section{Further details of simulation methods}
\label{app}

Our lattice model has energy function
\begin{eqnarray}
E =  \sum_{i,j}^\textrm{interactions} \epsilon_{C(i)-C(j)} +  \sum_{i}^\textrm{sites} \mu_{C(i)} \label{E}.
\end{eqnarray}
The first sum runs over all distinct nearest-neighbor interactions, and the second sum runs over all sites. $C(i)$ in \eqq{E} can be either w (white), b (blue) or r (red), depending on the color of node $i$; $\epsilon_{C(i)-C(j)}$ is the interaction energy between colors $C(i)$ and $C(j)$; and the chemical potential $\mu_{C(i)}$ is $\mu$, $-\ln(\fb^\textrm{s})$  and $-\ln(1-\fb^\textrm{s})$ for w, b and r, respectively. In the absence of pairwise energetic interactions (i.e. in notional `solution'), the likelihood that a given site will be white, blue or red is respectively $\left\{p_{\rm w}, p_{\rm b}, p_{\rm r} \right\} = \left\{{\rm e}^{-\mu},\fb^\textrm{s},1-\fb^\textrm{s}\right\} \left(1+{\rm e}^{-\mu} \right)^{-1}$.

Monte Carlo simulations were done as follows. We started with a simulation box that is 400 sites wide and 40 sites high, with the first six columns populated with the equilibrium checkerboard structure. We selected a node at random, and proposed a change of color of that node. If the chosen node was white, we attempted to make it colored; if the chosen node was colored, we attempted to make it white. If the chosen node was white, then we proposed to make it blue with probability $\fb^\textrm{s}$; otherwise, we proposed to make it red. No red-blue interchange was allowed, mimicking the idea that unbinding events are required in order to relax configurational degrees of freedom. To maintain detailed balance with respect to the stated energy function, the acceptance rates for these moves were as follows ($\Delta E$ is the energy change resulting from the proposed move):
\begin{eqnarray}
\textrm{r} \to \textrm{w} &:& \min(1,(1 - \fb^\textrm{s})\exp[-\Delta E]); \nonumber \\
\textrm{w} \to \textrm{r} &:& \min(1,(1 - \fb^\textrm{s})^{-1}\exp[-\Delta E]); \nonumber \\
\textrm{b} \to \textrm{w} &:& \min(1,\fb^\textrm{s} \exp[-\Delta E] ); \nonumber \\
\textrm{w} \to \textrm{b} &:& \min(1,(\fb^\textrm{s})^{-1} \exp[-\Delta E]  ). \nonumber
\end{eqnarray}
The notional solution abundances of red and blue are controlled by the chemical potential term that appears in \eqq{E} and therefore in the term $\Delta E$. Our choice to insert blue particles with likelihood $\fbs$ does not by itself result in a thermodynamic bias for one color over the other (because this bias in proposal rate is countered by the non-exponential factors in the acceptance rates). Instead, we bias insertions so that the dynamics of association is consistent with the thermodynamics of the model. For instance, if blue particles are more numerous in solution than red ones, we consider it to be physically appropriate to insert blue particles into the simulation box more frequently than red particles. Consider the limit of large positive $\mu$: the `solid solution' that results as the box fills irreversibly with colored particles will have a red:blue stoichiometry equal to that of the notional solution only if blue particles are inserted with likelihood $\fbs$. (As a technical note, the chemical potential term present in $\Delta E$ ends up simply canceling the non-exponential factors in the acceptance rates, but we have chosen to write acceptance rates as shown in order to make clear which pieces are imposed by thermodynamics, and which pieces we have chosen for dynamical reasons).

We also imposed a kinetic constraint that prevents any change of state of a lattice site that possesses only colored neighbors. This constraint, which respects detailed balance, is intended to model the fact that relaxation dynamics within solid structures is slow. In some simulations we omit the kinetic constraint in order to assess the outcome of slow internal evolution on timescales longer than we could otherwise access. In some regimes such constraint-free evolution leads rapidly to equilibrium, while in others it does not. For instance, for the inter-component interaction energies used to obtain \f{Fig:Crocker2}, grown structures evolve quickly to equilibrium if the kinetic constraint is not used. The kinetic constraint is therefore needed in order to capture the physical character of growth seen in experiments. By contrast, for the interaction energies used to obtain \f{FigFbs}, grown structures evolved even in the absence of the kinetic constraint fail to reach equilibrium, because of the deep kinetic traps associated with interaction energies large on the sale of $\kt$. There we can omit the kinetic constraint in order to effectively simulate longer, and still obtain nontrivial results. 

The parameter values $(\epsilon_\textrm{bb},\epsilon_\textrm{br},\epsilon_\textrm{rr})$ obtained from Refs.\c{Whitelam2014a},\c{Kim2009} and\c{Sue2015} and marked on \f{FigGrowthProtocol} are $(-3.5,-2,-3.5)$, $(-4.0,-3.21,-2.8)$ (also see \Tab{Crocker1}), and ($70$,$-7$,$0$).

Inherent structures in $d=\{1,...,6\}$ used to make \Fig{fig_flory}b and \Fig{fig_dimension} were obtained using zero-temperature single-spin-flip moves~\cite{Sue2015} (also see \Fig{Fig0}) starting from initial conditions in which all $\sim$2000 sites of the periodic system are randomly colored red or blue, with equal likelihood (\Fig{Fig0}). This procedure was carried out until no more spin flips occurred. At least 100 independent inherent structures were obtained for each datapoint. As shown in \Fig{Fig0}d, the average value of the resulting $\fb$ is insensitive to system size.

\begin{acknowledgments}
This work was done at the Molecular Foundry at Lawrence Berkeley National Laboratory, supported by the Office of Science, Office of Basic Energy Sciences, of the U.S. Department of Energy under Contract No. DE-AC02-05CH11231. 
\end{acknowledgments}

%

\clearpage

\linespread{1.5}
\setcounter{figure}{0}
\setcounter{equation}{0}
\setcounter{page}{1}
\makeatletter 
\renewcommand{\thefigure}{S\@arabic\c@figure}
\renewcommand{\thesection}{S\@arabic\c@section}
\renewcommand{\theequation}{S\@arabic\c@equation}
\renewcommand{\thetable}{S\@arabic\c@table}
\renewcommand{\thepage}{S\@arabic\c@page}
\makeatother

\makeatletter
\g@addto@macro\@floatboxreset\centering
\makeatother

\linespread{1}

\onecolumngrid
\appendix

~\vfil
\begin{center}
{\Huge Supplementary Materials}
\end{center}
\vfil
\begin{figure*}[h!]
  \centering
  \includegraphics[width=0.5\textwidth]{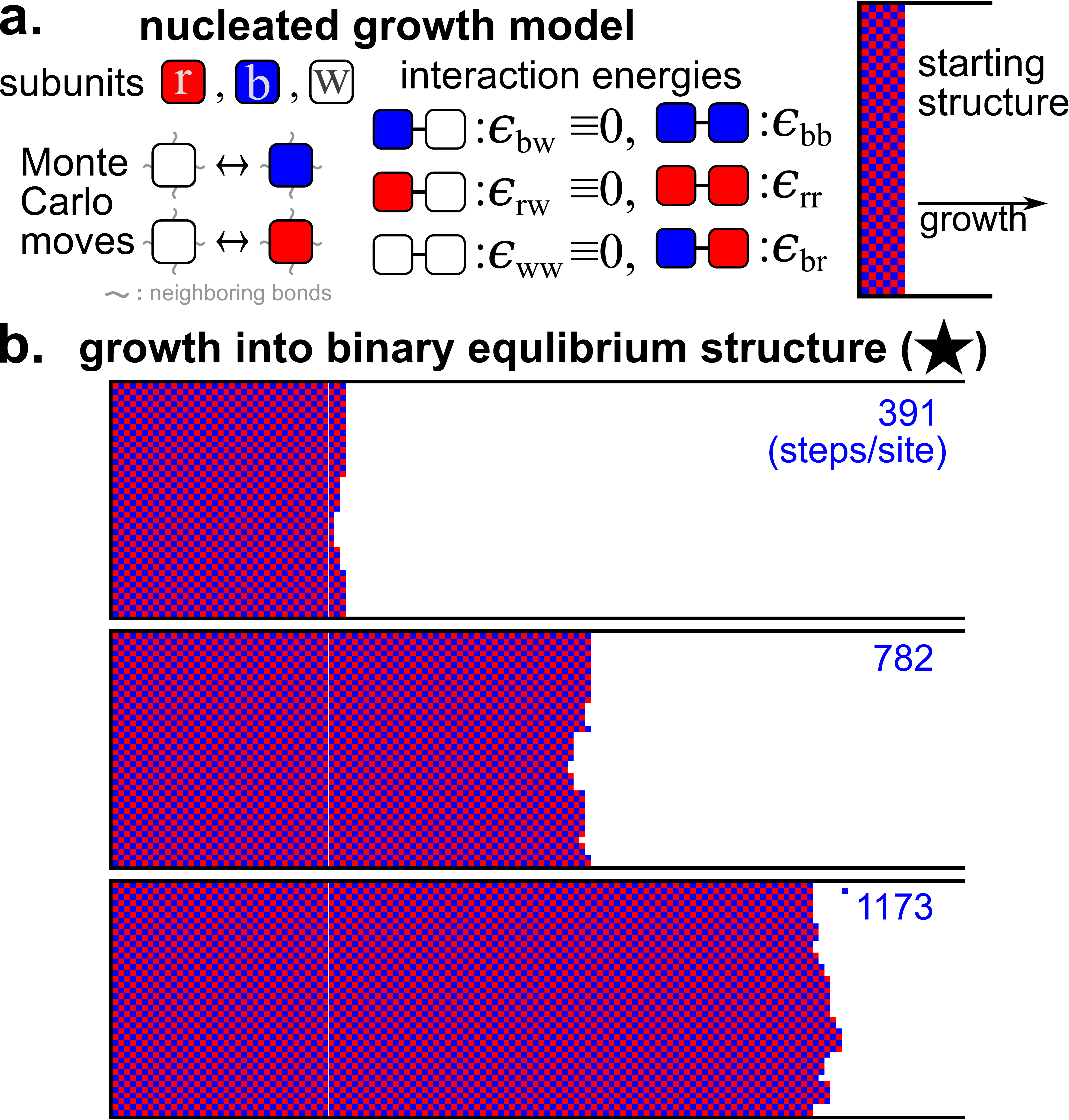}
  \caption{An example trajectory from the lattice model of growth described in the main text, showing the emergence of a structure very similar to the equilibrium checkerboard one. The long edges of the simulation box are periodic while the short edges are not, and growth occurs from the `seed' planted at the left-hand corner of the box. \label{FigGrowthProtocolExample}}
\end{figure*}
\vfil

\clearpage

~
\vfil

\begin{table*}[h]
\begin{tabularx}{1.0\textwidth}{|| Y || Y | Y || Y | Y | Y || Y | Y | Y ||}
\multicolumn{1}{c}{~} & \multicolumn{2}{c}{Parameters from \cc{Kim2009}} & \multicolumn{3}{c}{Energies from \cc{Kim2009}} & \multicolumn{3}{c}{Our parameter equivalents} \\
\hline 
 & $\Delta\Delta G$ & $\alpha$ & $|E_\textrm{b}^\textrm{AA}|$ & $|E_\textrm{b}^\textrm{AB}|$ & $|E_\textrm{b}^\textrm{BB}|$  & $\epsilon_\textrm{bb}$  & $\epsilon_\textrm{br}$  & $\epsilon_\textrm{rr}$     \\
hierarchy \# & ~                & $\equiv\textrm{e}^{\Delta\Delta G/\kt}$ &                          & $=|E_\textrm{b}^\textrm{AA}|/\alpha$ & $=|E_\textrm{b}^\textrm{AA}|/\alpha^2$ & =$-|E_\textrm{b}^\textrm{AA}|$ & =$-|E_\textrm{b}^\textrm{AB}|$ & =$-|E_\textrm{b}^\textrm{BB}|$ \\
\hline
$1$ & $1.25 $ $\kt$ & $3.50$ & $4 $ $\kt$ & $1.15 $ $\kt$  & $0.33 $ $\kt$ & $-4 $ $\kt$ & $-1.15 $ $\kt$  & $-0.33 $ $\kt$  \\
\hline
$2$ &  $0.22$ $\kt$             & $1.25$                                      & $4$ $\kt$                   & $3.21$ $\kt$ & $2.58$ $\kt$  & $-4$ $\kt$   & $-3.21 $ $\kt$ & $-2.58 $ $\kt$ \\
\hline 
\end{tabularx}
\caption{Interaction energy parameters of the binary nanoparticle mixture used in Ref.\c{Kim2009}. We take the A (resp. B) component of that reference to be the blue (resp. red) component of our model. Both energetic hierarchies used in Ref.\c{Kim2009} result in the hierarchy $\epsilon_{\rm bb} <\epsilon_{\rm br} <\epsilon_{\rm rr}$ in our language. The first hierarchy (characterized by $\Delta\Delta G=0.22$ $\kt$) resulted in the incorporation of red impurities in a majority blue structure, while the second hierarchy ($\Delta\Delta G=1.25$ $\kt$) does not. In \cc{Kim2009} the first energetic hierarchy resulted in the incorporation into blue structures of red impurity fraction 0.092$\pm$0.009 for a 50:50 blue:red solution stoichiometry, and a red impurity fraction of 0.0154$\pm$0.0025 for a 90:10 stoichiometry.}
\label{Crocker1}
\end{table*}

~\vfil
\begin{figure*}[h!]
\centering
  \includegraphics[width=0.8\textwidth]{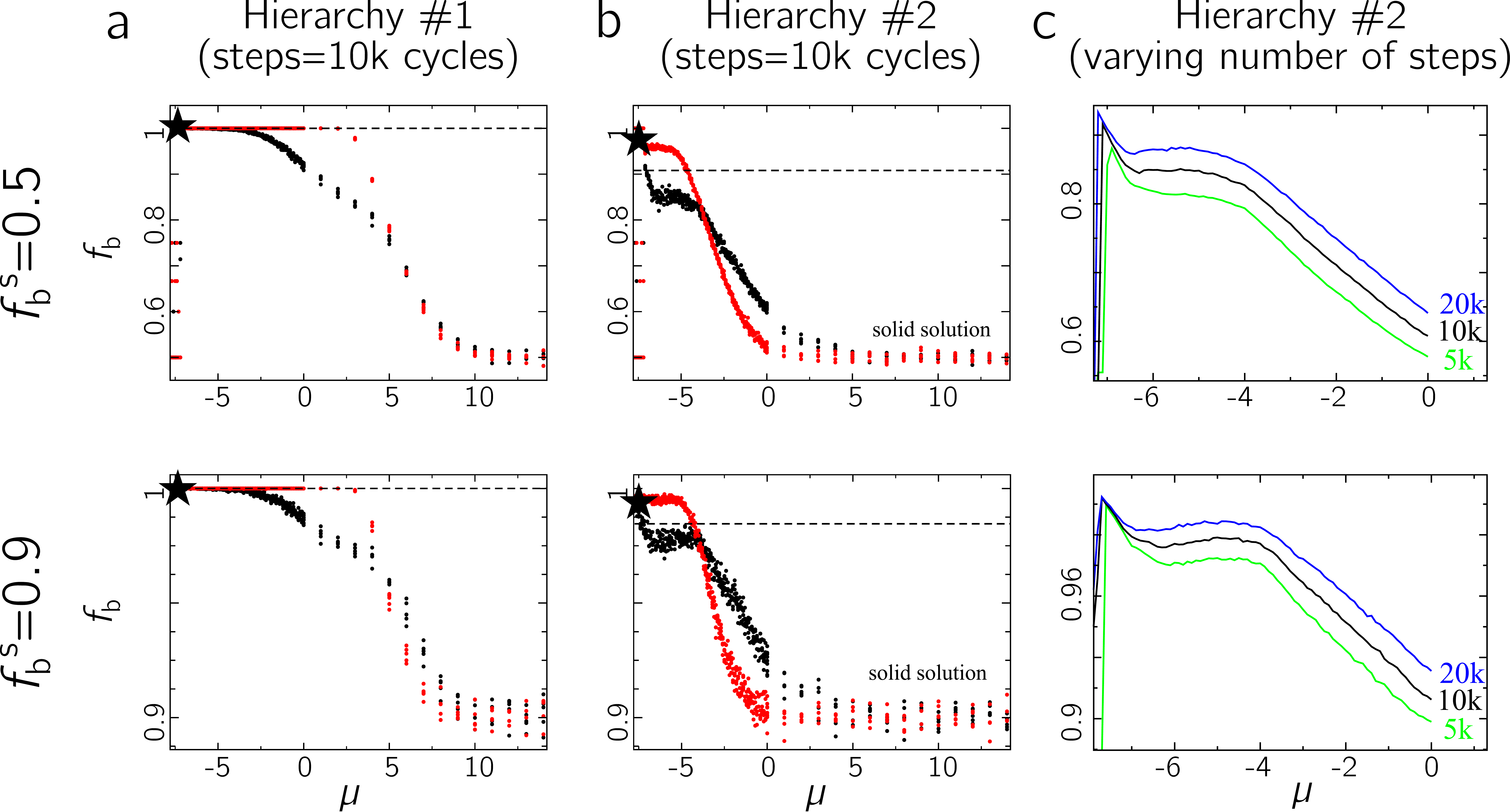} 
  \caption{Our model reproduces qualitatively the nonequilibrium behavior seen in the growth experiments of \cc{Kim2009}. Results shown here are from growth simulations with (red) and without (black) the kinetic constraint. The fraction of colored components in notional solution that are blue is 0.5 (top) and 0.9 (bottom). We used the red-blue interaction energies obtained from \Tab{Crocker1}, with \#1 used in panel (a) and \#2 used in panels (b,c). Two aspects of the experiments of Kim {\it et al.}\c{Kim2009} are reproduced by our simulations in the presence of the kinetic constraint. 1) Kim {\it et al.} reported the absence of red impurities in structures grown slowly using the energy hierarchy \#1; the same is true in our simulations for a range of low growth rates (in panel (a), dashed lines are the experimental value from \c{Kim2009}). 2) Kim {\it et al.} reported the presence of a plateau in the `purity fraction' $\fb$ as a function of growth rate, between near-equilibrium and far-from-equilibrium growth regimes using energy hierarchy \#2. A plateau is also seen in our simulations (panel b; we do not expect the numerical value of this plateau to match that seen in \cc{Kim2009}, because our lattice model has a geometry different to that of the fcc crystals grown in experiment). In panel (c) we show that structures allowed to evolve post-growth evolve slowly towards equilibrium.
\label{Fig:Crocker2}}
\end{figure*}
\vfil

\clearpage

\begin{figure*}[t!]
  \centering
  \includegraphics[width=1.0\textwidth]{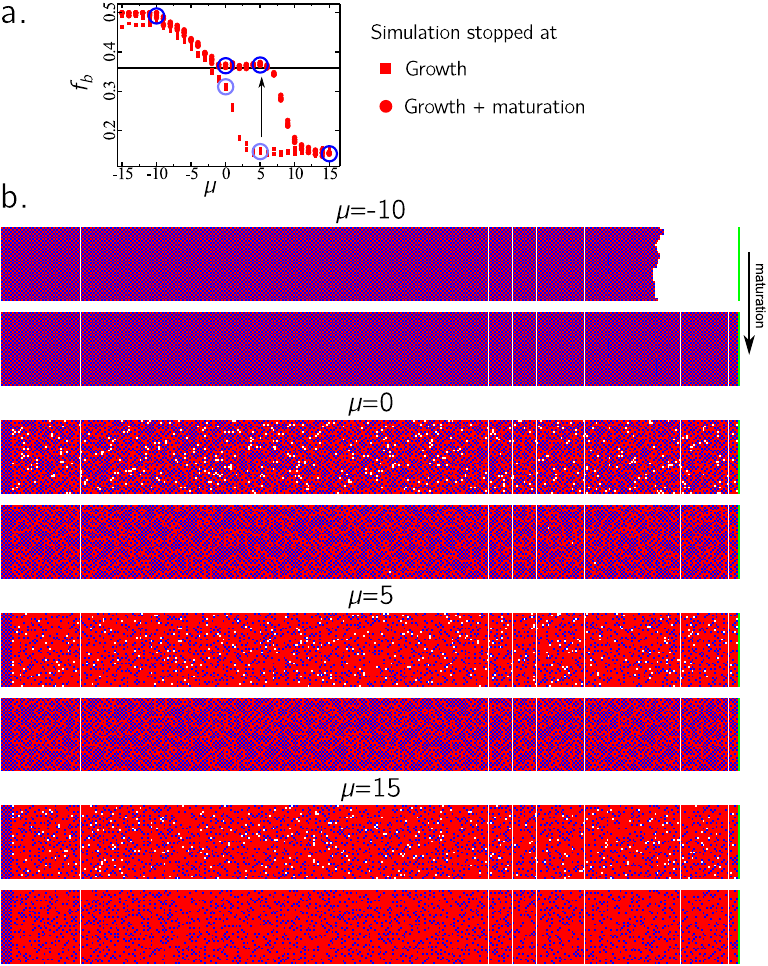}
  \caption{Panel (a) describes outcomes of growth (squares) and growth-plus-maturation (circles) for solution blue fraction $\fb^\textrm{s}=.2$ and $\epsilon_\textrm{br}<0\equiv\epsilon_\textrm{rr}\ll\epsilon_\textrm{bb}$ (\f{FigGrowthVsMaturation}).
  The snapshots in Panel (b) (top: growth; bottom: growth plus maturation) are taken from circled points in Panel (a). The maturation simulations were stopped at 10,000 cycles or steps per site (using other values do not qualitatively change our results; \f{ManyCycles}).
\label{FigGrowthVsMaturationSnapshotsAll}}
\end{figure*}

\clearpage
~\vfil
\begin{figure*}[h!]
  \centering
  \includegraphics[width=0.8\textwidth]{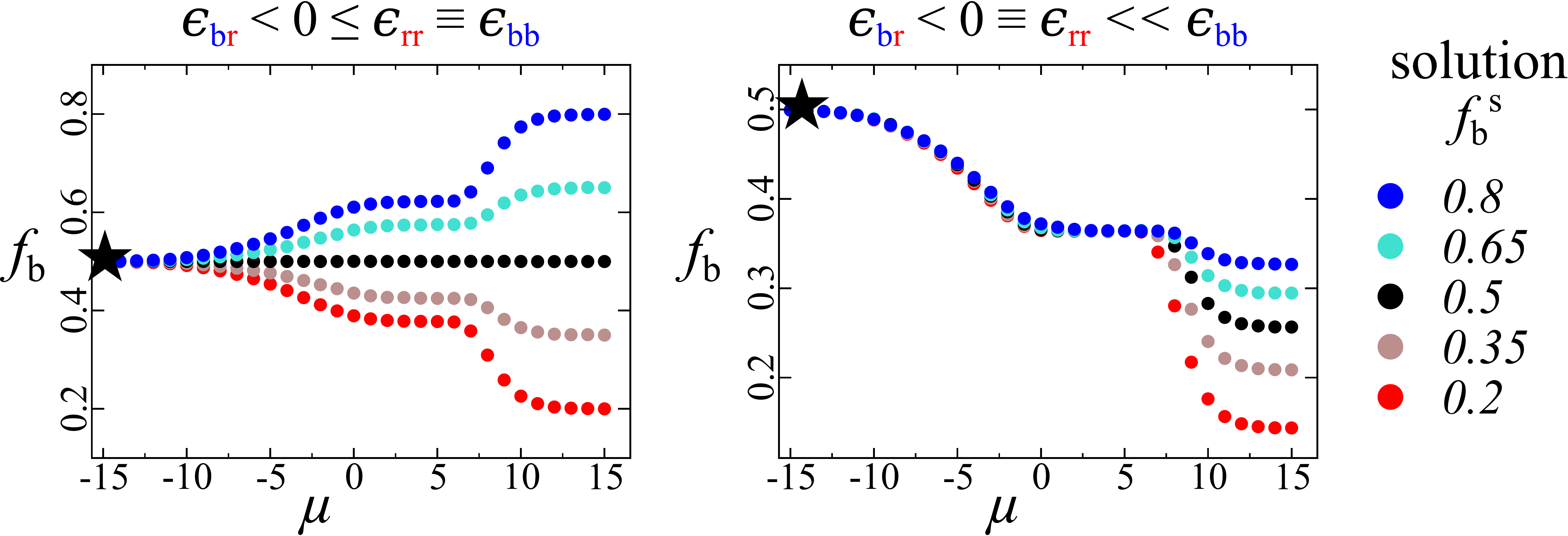}
  \caption{Maturation simulations performed at a range of solution blue fractions $\fb^\textrm{s}$. This figure is an extension of the lower panel (b) of \Fig{FigFbs}, and shows that `magic number' plateaux are insensitive to growth rate and solution concentration over a wide range of those parameters.
\label{MoreConditions}}
\end{figure*}

\vfil

\begin{figure*}[h!]
  \centering
  \includegraphics[width=0.95\textwidth]{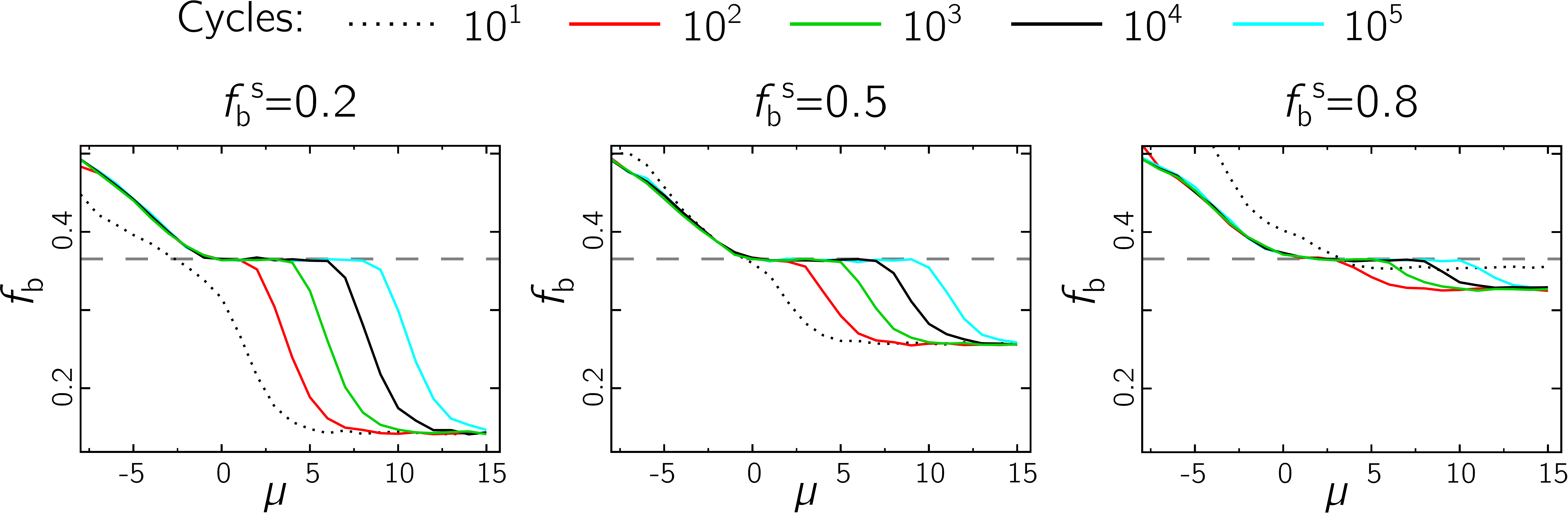}
  \caption{Maturation simulations were run with different stopping times, ranging from $10^1$ to $10^5$ cycles or steps per site. Except for the shortest stopping time (10 cycles; dotted line) all simulations shows plateaux at the expected magic number ratio (dashed gray line), with only the right end of the plateau being extended. $10^4$ is the maturation time utilized in all other figures within the manuscript.
\label{ManyCycles}}
\end{figure*}

\vfil

\clearpage
~
\vfil
\begin{figure*}[h!]
  \centering
  \includegraphics[width=0.8\textwidth]{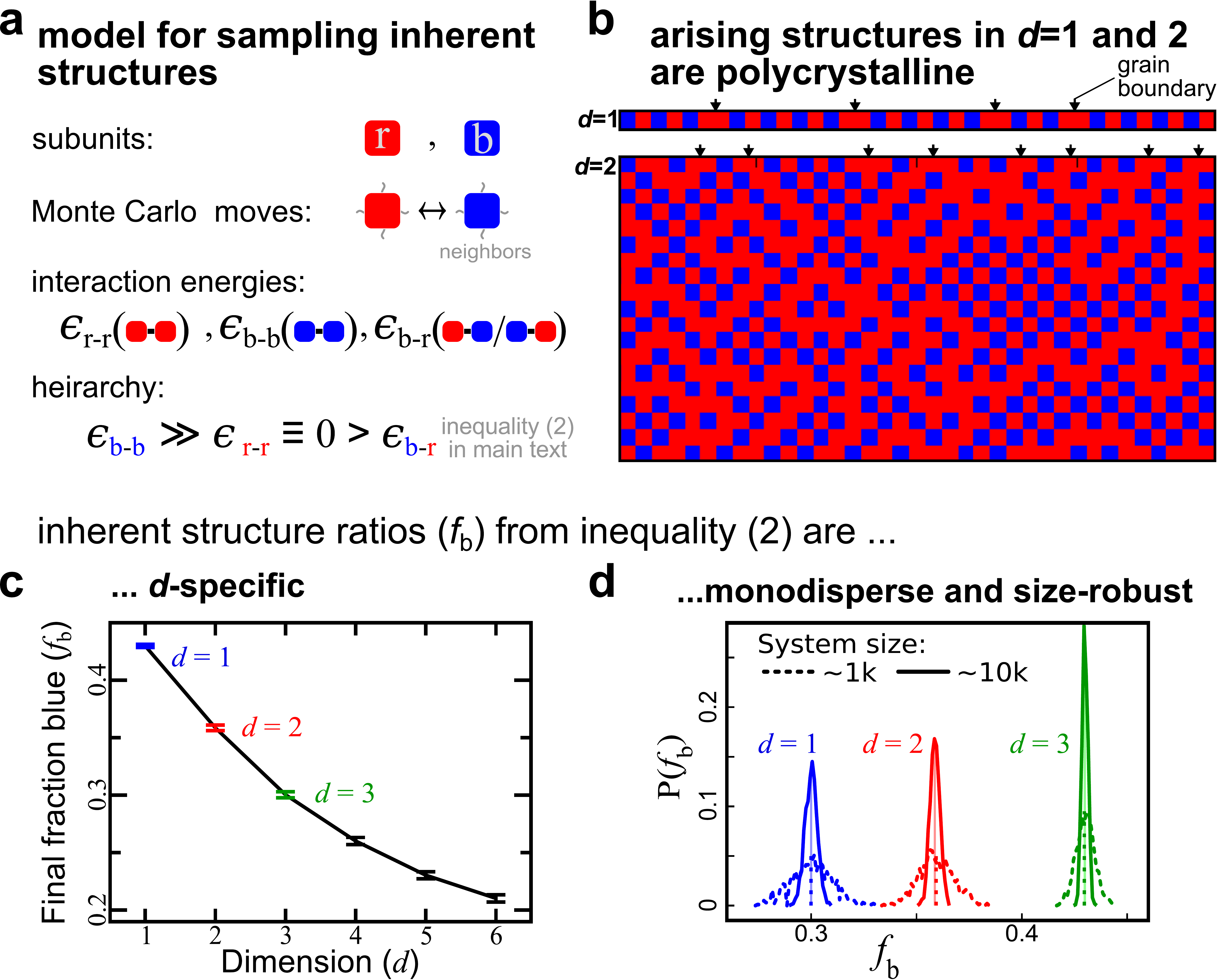}
  \caption{Utilizing a sampling algorithm (a; See Ref.~\citenum{Sue2015}) and the energy hierarchy $\epsilon_\textrm{br}<\epsilon_\textrm{rr}\equiv0\ll\epsilon_\textrm{bb}$ ensures that all randomly populated conformations on a lattice relax to kinetically-trapped structures (or inherent structures) that belong to a class of binary polycrystals. These polycrystals, shown in (b) for $d=1$ and $2$, are marked by binary (alternating red-blue) domains separated by all-red boundaries. Interestingly, each inherent structure describes topology-specific composition (c; $\fb$ is the fraction of blues in the inherent structure) that is irrespective of system size and solution concentration (d). Interestingly, these polycrystalline structures appear to be ``universally'' accessible via growth mechanisms\cite{Sue2015} as well as nucleated growth simulations (\Fig{FigGrowthVsMaturation}b).
\label{Fig0}}
\end{figure*}
\vfil

\clearpage

\vfil
\begin{figure*}[h!]
\centering
\includegraphics[width=1.0\textwidth]{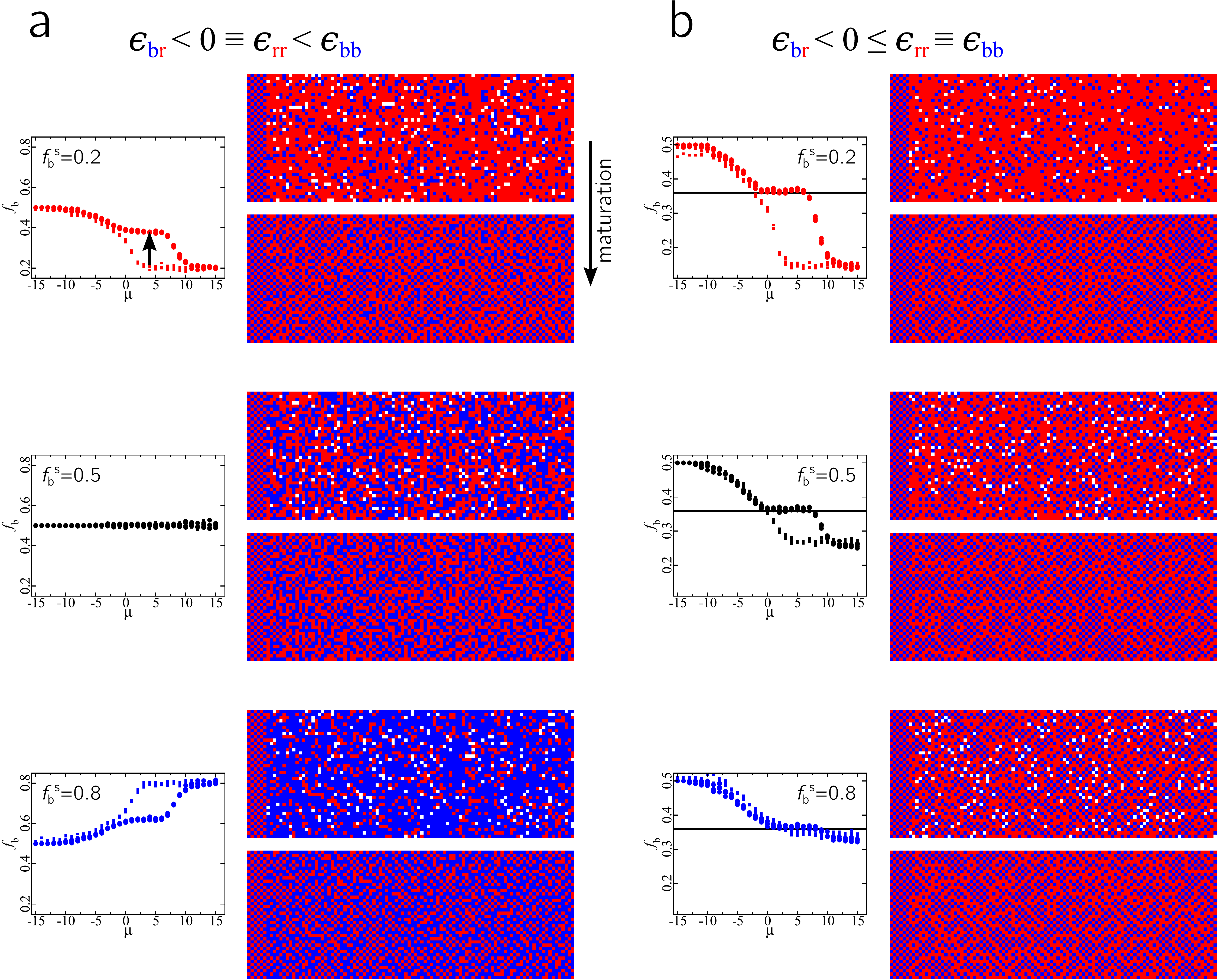}
  \caption{Growth (squares) versus growth+maturation (circles). Each graph displays the relationship between growth rates and outcomes from growth (squares) and growth followed by maturation (circles) for various solution concentrations ($\fb^\textrm{s}\in\{$\textcolor{red}{.2},.5,\textcolor{blue}{.8}$\}$; from top to bottom) and energy hierarchies--$\epsilon_\textrm{br}<0\equiv \epsilon_\textrm{rr}\equiv\epsilon_\textrm{bb}$ (a) and $\epsilon_\textrm{br}<\epsilon_\textrm{rr}\equiv0\ll\epsilon_\textrm{bb}$ (b). The lines in the graphs in (b) represent the magic number ratios obtained upon maturation. To the right of each graph is one snapshot of growth and one snapshot of growth+maturation at $\mu=4$.
\label{FigGrowthVsMaturationAll}}
\end{figure*}
\vfil

~
\vfil

\begin{figure*}[h!]
  \centering
  \includegraphics[width=1.0\textwidth]{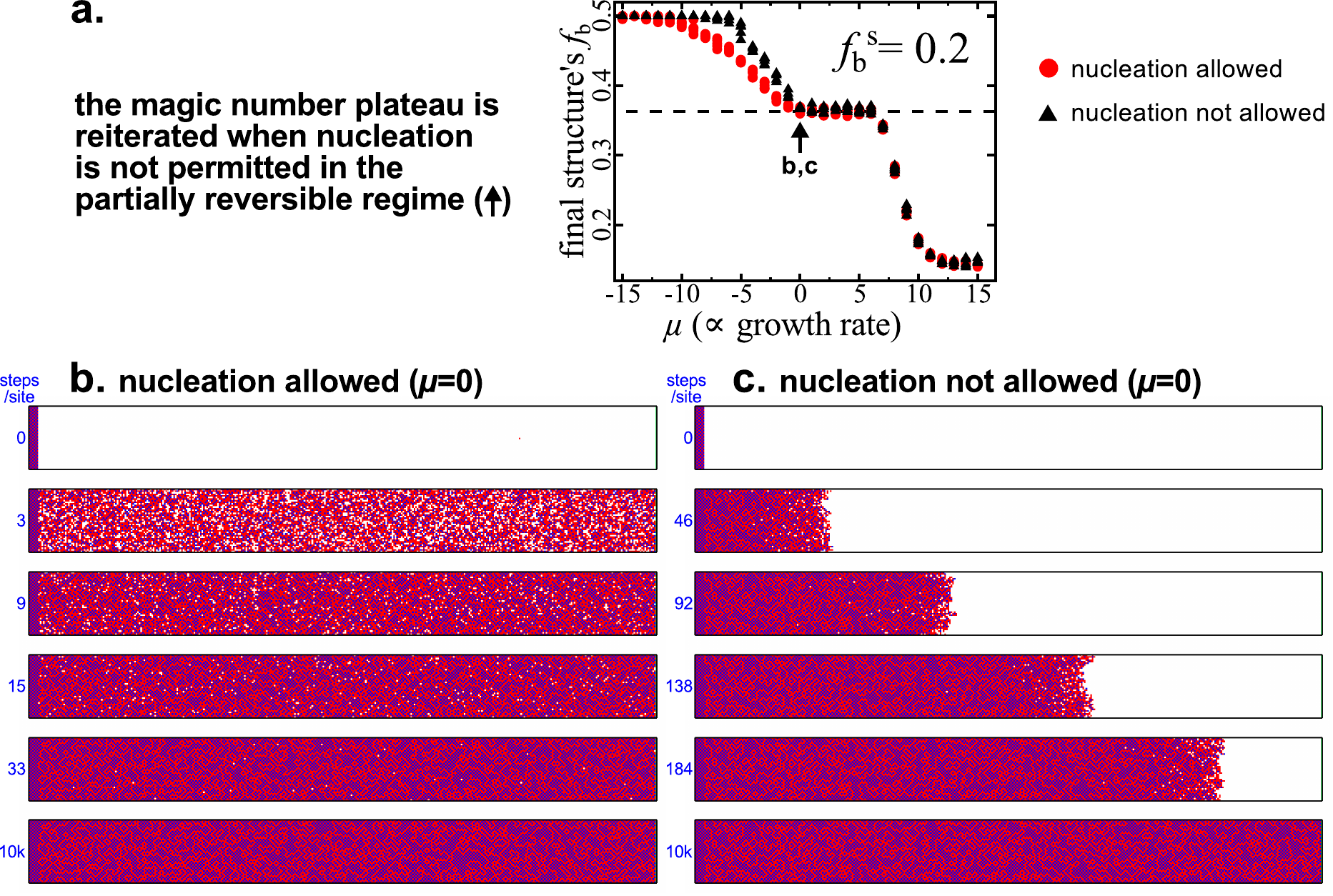}
  \caption{Growth with and without nucleation results in similar final `magic number' outcomes. This is partially because the assembly close to the growth face (c) retains a low-occupancy region that resembles the bulk of the multi-nucleated assembly (b). Maturation of these porous regions occur either after growth (b) or during growth (c) to form the polycrystalline structure with the characteristic magic number ratio.
\label{FigNucleation}}
\end{figure*}

\end{document}